\documentclass[runningheads]{llncs}
\usepackage{graphicx}
\usepackage{float}
\usepackage{cite}

\begin{document}
\title{Compile-time Parallelization of Subscripted Subscript Patterns
}
%
%
\author{Akshay Bhosale and Rudolf Eigenmann}
\authorrunning{F. Author et al.}
%
\institute{ Univeristy of Delaware, Newark, DE, USA.\\
  \email{\{akshay,eigenman\}@udel.edu}
}
\maketitle              
\begin{abstract}
An increasing number of scientific applications are making use of irregular data access patterns. An important class of such patterns involve subscripted subscripts, wherein an array value appears in the index expression of another array. Even though the information required to parallelize loops with such patterns is often available in the program, present compiler techniques fall short of analyzing that information. In this paper we present a study of subscripted subscripts, the properties that define the subscript arrays, and an algorithm based on {\em symbolic range aggregation}, that will help prove the presence of some of the properties of the subscript array in the program. We show that, in an important class of programs, the algorithm can boost the performance from essentially sequential execution to close to fully parallel.

\keywords{automatic parallelization \and subscript array \and aggregation}
\end{abstract}
\section{Introduction}
\label{sec:intro}

In this paper we develop compile-time analysis techniques to understand
program patterns that involve subscripted subscripts, such as a[b[i]], and to use this
information for automatic parallelization. To the best of our knowledge,
this is the first time compile--time techniques have been developed that can
successfully parallelize programs exhibiting subscripted subscripts without the
help of runtime techniques. In this initial paper, we have developed the
algorithms, and applied them by hand. The full implementation will be described
in a forthcoming contribution.

Two motivations met to initiate the research behind this paper. Scientific
applications increasingly make use of irregular data structures,  often
represented in the form of sparse matrices. Algorithms operating on such data
structures commonly involve indirection arrays, leading to subscripted
subscript patterns. Currently, there are no
known {\em compile-time} analysis techniques that can automatically parallelize
such patterns. The second motivation came from an effort to find loops in
existing scientific programs that were parallelized manually, but, when feeding
the serial versions of the programs to the Cetus~\cite{BMLA12} translator, it
could not detect that the loops were parallel. We did this for the latest
version of the NAS parallel benchmarks. We found that (other than a few cases
that exhibited compiler bugs) the primary reason Cetus couldn't match manual
parallelization was the presence of subscripted subscripts. Similar experiments with other compilers or translators, such as Rose~\cite{quinlan2011rose}, Intel's ICC Compiler~\cite{IntelICC} and the PGI~\cite{PGI2018} compiler also yielded the same result. We also found that,
in many such programs, {\em the necessary and sufficient information that the
involved loops can in fact be parallelized was present in the program code and
was not dependent on the input data.} While investigating this information can be
complex,  the opportunity {\em exists} to develop compile-time analyses to do
  this detection. The present contribution pursues this goal.

The key to successful compile-time parallelization of the subscripted subscript
patterns we encountered is the detection of certain properties of the
indirection arrays and their content, such as monotonicity and
injectivity. This observation, per se, is not new. Others have developed
techniques that make use of run-time methods, such as inspector/executor
schemes to detect the properties and then execute the involved loops in
parallel~\cite{MYCD19}, or they have successfully analyzed certain properties,
but  not yet succeeded in  automatic
parallelization~\cite{LiPa00 , mohammadi2018extending}. In our work, we
found that the needed properties can be derived from the code that creates and
modifies the subscript array contents.

In this paper, we present a first analysis to do so,
which is sufficient to parallelize a class of programs. We also present a
study of other patterns, for which the development of analysis techniques is
our ongoing work. 

In summary, in this paper we make the following contributions: 
\begin{itemize}
\item We present an empirical study of subscript patterns and argue that these patterns can be investigated at compile time, in a way
  that enables automatic parallelization.
\item We describe an algorithm that analyzes the array properties that are
  needed for automatic parallelization of an initial program viz. representative of a class of programs.
\item We present performance results after applying the techniques and
  automatic parallelization by hand.
\end{itemize}

The remainder of the paper is organized as follows. Section~\ref{sec:study}
presents our study of subscripted subscript
patterns. Section~\ref{sec:algorithm} describes the algorithm for deriving
properties of arrays that show up as index arrays in to-be-parallelized
loops. Section~\ref{sec:related} discusses related work. Section~\ref{sec:eval} shows performance results of programs
exhibiting subscripted subscript patterns, after applying the techniques
presented in Section~\ref{sec:algorithm}. Section~\ref{sec:conclude} presents conclusions.

\section{Empirical Study of Subscripted Subscript Patterns}
\label{sec:study}

\begin{figure}[H]
\centering\includegraphics[width=4.5in]{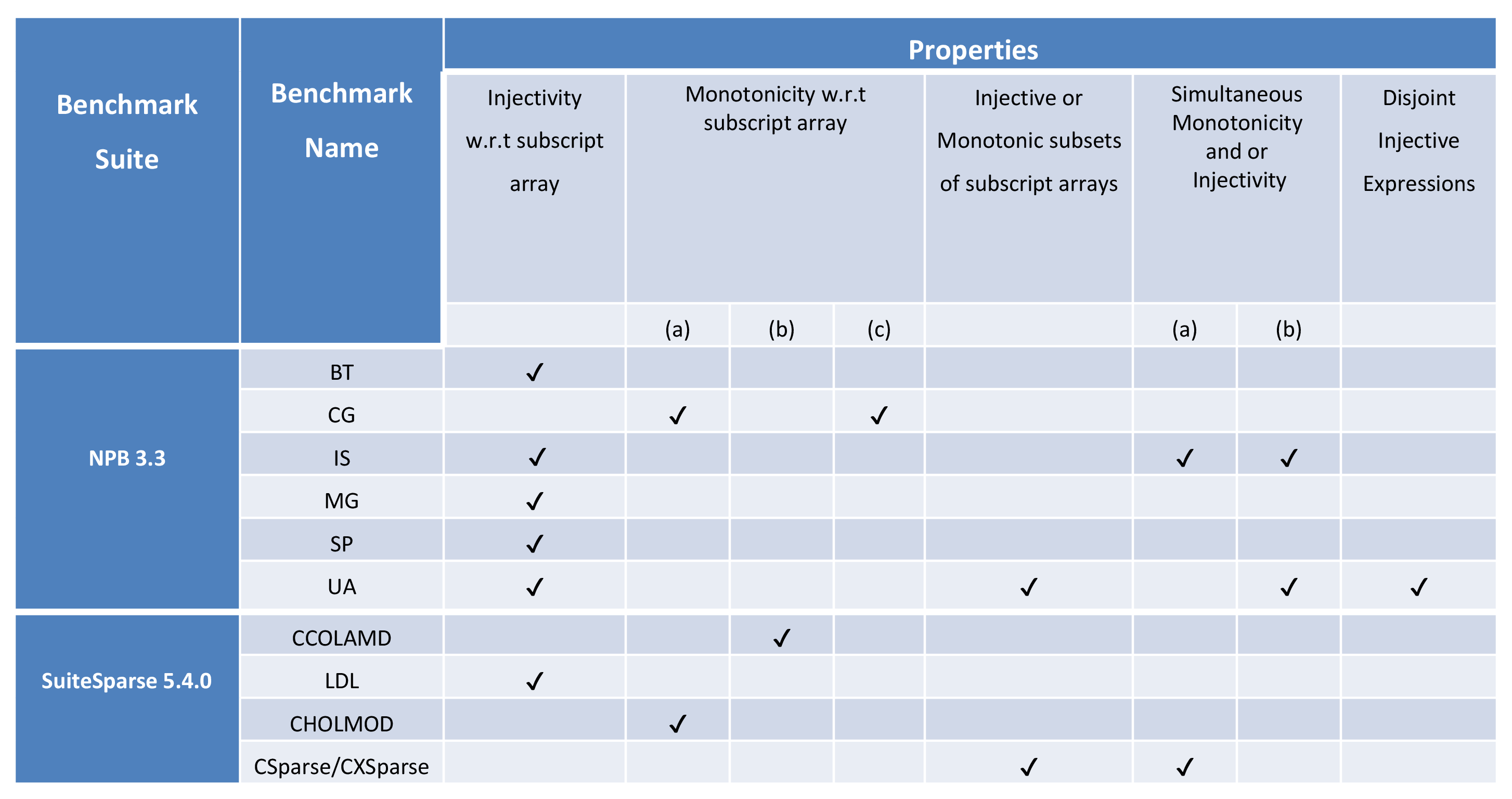}
\caption{Analysis of subscripted subscript patterns in the NAS parallel benchmarks v3.3.1. and the SuiteSparse v5.4.0. } 
\label{fig:Properties-table1}
\end{figure}

Figure~\ref{fig:Properties-table1} shows our findings of the use of index arrays in programs of the most recent NAS Parallel Benchmark Suite~\cite{bailey2011parallel} and the SuiteSparse~\cite{doi:10.1137/1.9780898718881} programs. We analyzed the programs for loops with subscripted subscript patterns and properties of the index arrays that make the loops parallelizable. 

\vspace{-5mm}
In the NPB Suite, of the ten programs, we found six that contained parallel loops accessing arrays that used the values of other arrays in their subscript expressions. In the SuiteSparse, four of the eight programs analyzed contained parallelizable loops with such patterns. Common to all these patterns is that the parallel loop contains a single array write reference with a subscript expression that contains the value of another array, but there is no read reference of the written array. To parallelize the loop, the compiler would need to prove that there is no self output dependence. The properties are:

\vspace{-2mm}
\begin{figure}[H]
\footnotesize
\begin{verbatim}
From – UA benchmark ( NPB 3.3 ):
1: for (miel = 0; miel < nelt; miel++) {
2:    iel = mt_to_id[miel];
3:    id_to_mt[iel] = miel;
4:  }

\end{verbatim}
\vspace{-5mm}
\caption{The outer loop is parallelizable, as array $mt\_to\_id$, whose value is used in the subscript of array $id\_to\_mt$, is injective. Injectivity of $mt\_to\_id$ ensures that writes of $id\_to\_mt$ go to different elements in different iterations. Hence there is no output-dependence.}
\label{fig:fig1}
\end{figure}
\vspace{-6mm}

\begin{enumerate}
    \item Injectivity: An array is said to be injective if $a[i] \neq a[j]$, $\forall i \neq j.$ Write accesses to an array containing an injective array as a subscript can be done in parallel, because all array accesses reference different elements. Figure~\ref{fig:fig1} shows an example where injectivity is the defining property for parallelization.

    \vspace{2mm}
    \item Monotonicity: We encountered three variations of montonicity in the benchmark applications.
 
\begin{enumerate}

    \vspace{2mm}
    \item Montonically increasing or decreasing: An array is monotonically increasing if $a[i] \leq a[j]$, $ \forall i < j $ and monotonically decreasing if $a[i] \geq a[j]$,  $\forall i < j$. This implies non-strict monotonicity. Figure~\ref{fig:fig2a} shows an example where non-strict monotonicity is sufficient for parallelizing the outer, `j' loop.
 
\vspace{-2mm}   
\begin{figure}[H]
\footnotesize
\begin{verbatim}
From – CG benchmark ( NPB 3.3 ):
1:  for (j = 0; j < lastrow - firstrow + 1; j++) {
2:     for (k = rowstr[j]; k < rowstr[j+1]; k++) {
3:       colidx[k] = colidx[k] - firstcol;
4:    }
5: }
\end{verbatim}
\caption{In this example, values of array $rowstr$ are used in the subscript of array $colidx$. Since $rowstr$ is monotonic, the ranges $rowstr[j]$ to $rowstr[j+1]-1$ do not overlap across iterations of the outer loop.}
\label{fig:fig2a}
\end{figure}

  \vspace{-4mm}
    \item Strictly monotonically increasing or decreasing: An array is strictly monotonically increasing if $a[i] < a[j]$, $ \forall i < j $ and strictly monotonically decreasing if $a[i] > a[j]$, $\forall i < j$. Strict monotonicity implies injectivity.

    \vspace{2mm}
    \item Monotonic difference between arrays: In this case, the difference between two arrays is monotonic and the loop index traverses this difference. Figure~\ref{fig:fig2b} shows an example. Non-strict monotonicity is sufficient for parallelization of the loop in this case.
\end{enumerate}

\vspace{-2mm}
\begin{figure}[H]
\footnotesize
  \begin{verbatim}
From – CG benchmark ( NPB 3.3 ):
1: for ( j = 0; j < nrows; j++){
2:    if ( j > 0){
3:    j1 = rowstr[j] - nzloc[j-1];
4:   } 
5:    else{
6:     j1 = 0;
7:   }
8:   j2 = rowstr[j+1] - nzloc[j];
9:   nza = rowstr[j];
10:  for( k = j1; k < j2; k++){
11:      a[k] = v[nza];
12:      colidx[k] = iv[nza];
13:      nza = nza + 1;
14:   }
15: }
\end{verbatim}
\caption{In this example, the difference between consecutive elements of arrays $rowstr$ and $nzloc$ is monotonic. The range from $j1$ to $j2-1$ is non-overlapping across iterations of the `j' loop, which makes it parallelizable.}
\label{fig:fig2b}
\end{figure}

    \vspace{-5mm}
    \item Injective or Monotonic subsets: In this case only a part, or a subset, of the subscript array is monotonic (Non-strict or Strict) or injective. The loop can be parallelized, if the access pattern ensures that only those array elements are referenced that exhibit the property. In the benchmark codes that we analyzed we came across subscript arrays with injective subsets. Figure~\ref{fig:fig3} shows such an example.

\vspace{-2mm}  
\begin{figure}[H]
\footnotesize
 \begin{verbatim}
From CSparse ( SuiteSparse 5.4.0 )
1: for (i = 0; i < m; i++) {
2:   if (jmatch [i] >= 0) {
3:      imatch [jmatch [i]] = i;
4:    }
5:  }
\end{verbatim}  
\caption{In this example the subset formed by only the non-negative elements of array $jmatch$ is injective. The access pattern makes sure that elements from only this injective subset are referenced.}
\label{fig:fig3}
\end{figure}

   \vspace{-3mm}
   \item Simultaneous monotonicity and/or injectivity: Some benchmark applications involve loop nests with arrays consisting of multiple levels of indirection at their subscripts, such as A[B[C[i]]]. Both subscript arrays possess certain properties for the enclosing loop to be parallelizable.
   
   \vspace{2mm}
\begin{enumerate}
    \item Simultaneous Monotonicity and Injectivity: The innermost subscript array is monotonic whereas the outermost subscript array is injective  and hence the outer loop becomes parallelizable. Figure~\ref{fig:fig4} shows such an example, extracted from the SuiteSparse library.

    \vspace{2mm}
    \item Simultaneous Injectivity: Both  subscript expressions are injective.
    Figure~\ref{fig:fig5} shows an example. Injectivity of multiple nested subscript expressions ensures that different values of the innermost expression -- typically the loop index -- imply different values of the outermost expression. Note that, in addition to the subscript arrays, the expressions in which the arrays appear need to be injective as well. 
  
 \vspace{-2mm}     
\begin{figure}[H]
\footnotesize
\begin{verbatim}
From CSparse ( SuiteSparse 5.4.0 )
1:  for (b = 0; b < nb; b++){
2:    for (k = r[b]; k < r[b+1]; k++) {
3:         Blk[p[k]] = b;
4:     }
5:  }
\end{verbatim}
\caption{The outer loop in this example can be parallelized since
array $r$ is monotonically increasing and array $p$ is injective. }
\label{fig:fig4}
\end{figure}
    
    \vspace{-2mm}

    \end{enumerate}
  
\item Disjoint Injective Expressions: In 4.(b), the subscript arrays were part of a single expression. In some benchmark applications, we came across loops where more than one expression factored in at the subscript of an array. As shown in Figure~\ref{fig:fig6}, each individual subscript expression produces a set of unique values. In addition, the two sets of values are mutually exclusive. This makes sure that the loop index maps to a unique set of values and hence the mapping is injective.
\end{enumerate}

\vspace{-5mm}
\begin{figure}[H]
\footnotesize
 \begin{verbatim}
From UA benchmark( NPB 3.3 )
1: for( index = 0; index < num_refine; index++){
2:      miel = action[index];
3:      iel = mt_to_id_old[miel];
4:      nelt = nelttemp+(front[miel]-1)*7;
5:          ...;
6:          ...;
7:          ...;
8:          ...;
9:   for( i = 0; i < 7; i++){
10:      tree[nelt + i] = ntemp + (( i + 1) % 8);
11:   }
12:         ...;
13:         ...;
14: }
\end{verbatim}
\caption{The subscript of array $tree$ on line 10 will contain arrays $action$ and  $front$. Both are injective. The expressions on lines 4 and 10 produce injective values in every iteration of the outer loop. Also the expression on line 4 is strictly monotonic,   increasing by 7 in successive iterations of the outer loop.}
\label{fig:fig5}
\end{figure}

\vspace{-10mm}
\begin{figure}[H]
\footnotesize
 \begin{verbatim}
 
From UA benchmark( NPB 3.3 ):
1: for (miel = 0; miel < nelt; miel++) {
2:    iel = mt_to_id_old[miel];
3:    if (ich[iel] == 4) {
4:      ntemp = (front[miel]-1)*7;
5:      mielnew = miel + ntemp;
6:    } else {
7:      ntemp = front[miel]*7;
8:      mielnew = miel + ntemp;
9:    }
10:   mt_to_id[mielnew] = iel;
11:   ref_front_id[iel] = nelt + ntemp;
12:  }
\end{verbatim}
\caption{In this example, two subscript expressions can appear at the subscript of array $mt\_to\_id$ on line 10. The two expressions , one on line 5 where $ntemp$ is $(front[miel] - 1)*7$ and on line 8 where $ntemp$ is $front[miel]*7$ produce sets of values which are strictly monotonic as well as mutually exclusive. Hence unique values appear at the subscript of $mt\_to\_id$ on line 10 in every loop iteration.}
\label{fig:fig6}
\end{figure}
\vspace{-5mm}

In the examples discussed above, the properties were verified to be present and the loops parallelizable due to these properties. A key observation in our study was that, in many cases, the program code creating the content of the subscript arrays is present in the programs, and an advanced programmer is able to determine from that code that the arrays have the necessary properties discussed in the above examples. The code creating the properties was not shown. Figure~\ref{fig:CGcode} shows the essence of one of these code sections (lines 1--15), followed by the to-be-parallelized loop.

\vspace{-5mm}
\begin{figure}[H]
\footnotesize
\begin{verbatim}
1: for( i = 0; i < ROWLEN; i++){
2:    count = 0;
3:    for ( j = 0; j < COLUMNLEN; j++){
4:       if( a[i][j] != 0){
5:          count++;
6:     	    column_number[index++] = j;
7:    	    value[ind++] = a[i][j]; 
8:	     }
9:    }
10:   rowsize[i] = count;
11: }
12: rowptr[0] = 0;
13: for( i = 1; i < ROWLEN + 1; i++){
14:     rowptr[i] = rowptr[i-1] + rowsize[i-1];
15:  }
16:
17: #pragma omp parallel for private(j,j1)
18: for( i = 0; i < ROWLEN+1; i++){
19:     if( i == 0 ){
20:        j1 = i;
21:    }
22:    else{
23:        j1 = rowptr[i-1];
24:   }
25:    for( j = j1; j < rowptr[i]; j++){
26:        product_array[j] = value[j] * vector[j];
27:    }
28: }
29:

\end{verbatim}
\vspace{-5mm}
\caption{Example of parallel loop using index array (lines 17--28) and the code filling the index array (lines 1--15). Here, monotonicity of array $rowptr$ enables the parallelization of the outer loop on line 18.}
\label{fig:CGcode}
\end{figure}

\vspace{-8mm}
In order to parallelize the outer loop (lines 17--28 ), one must ensure that the index ranges of the inner loop do not overlap for different values of $i$. These ranges are essentially from $rowptr[i-1]$ to $rowptr[i]$. They do not overlap if $rowptr[i-1] <= rowptr[i]$, which is the same as saying $rowptr$ is monotonic. Non-strict monotonicity is sufficient, in this case. This property can be derived from the code on lines 1--15. Line 14 creates a monotonic relationship between two consecutive array elements, if the value $rowsize[i-1]$  is non-negative.

This is the case, as the code on lines 1--10 assigns a non-negative value, $count$, to each array element within the index range of interest. $Count$ is initialized to $0$ (line 2) and is only incremented (line 5).

In our study we have found several similar patterns, hence Figure~\ref{fig:CGcode} represents a class of programs using subscripted subscripts. The next section presents a compiler algorithm that analyzes this class.
While our study found the more complex examples described in this section as well, a key observation was that the subscript array properties that are needed to guarantee parallelizability could be derived from the program code itself and were not dependent on input data. The content of the index arrays are input dependent, but the key properties needed for parallelization tend to be invariant.

\section{Compile-time Algorithm for Index Array Analysis}
\label{sec:algorithm}

\subsection{Algorithm Overview}
Our algorithm determines the properties described in the previous section as follows. The algorithm proceeds in program order, analyzing the loops in each nest from inside out. At each loop level, it proceeds in two phases to analyze the values of the variables of interest. There are two types of variables of interest: integer scalars and integer arrays with simple subscripts. For the current algorithm, ``simple subscript" means index expressions of the form $i+k$, where $i$ is the loop variable and $k$ is a constant.

Phase~1 analyzes the loop body and determines the effect of one iteration on the variable's value. Phase~2 then aggregates this expression across all iterations, determining the effect of the entire loop on the scalar or array. After Phase~2, the loop is collapsed, that is, it is substituted by a set of expressions representing the effect of the loop. The algorithm then proceeds with the next outer loop.

\subsection{Representation}
Our representation of a variable value is a symbolic expression that may contain ranges of the form $[l:u]$, where the lower bound $l$ and upper bound $u$ are symbolic expressions. Values are {\em may} ranges, i.e., the actual value may be any one in the range.
\begin{quote}
$x: [lb:ub]$ -- (scalar) variable $x$ has possible value range between $lb$ and $ub$ at current program point.
\end{quote}
For arrays, the representation includes a subscript range. Subscript ranges are {\em must} ranges, i.e., the indicated value holds for all array elements in that range. 

\begin{quote}
    $y: [sl:su] , [vl:vu]$ -- all elements of array $y$ in index range $sl:su$ have a value in the range of $vl$ to $vu$.
\end{quote} 

The representation makes use of several special symbols, referring to particular values of a variable being analyzed:
\begin{itemize}
    \item $\lambda$ refers to the value of the variable at the beginning of the loop iteration being analyzed. This will be used in Phase~1 and for scalars.
    \item $\Lambda$ refers to the value of the variable at the beginning of the loop. This is useful in the aggregation step of Phase~2. $\Lambda$ will also be used in the expression that represents the effect of the loop after collapsing, where it refers to the value before the expression.
    \item $\bot$ indicates an unknown value, e.g., if an expression is too complex for the compiler to analyze or represent.
    \item Instead of a value (range) the representation may indicate an array property, such as Monotonic\_inc/dec or Injective.
\end{itemize}

\subsection{Algorithm for Phase 1}

Our algorithm makes use of the symbolic range analysis~\cite{BlEi95} method to analyze the body of the given loop. 
Both integer scalars and arrays are analyzed. Scalar values are initialized to $\lambda$. Phase~1 computes a symbolic range expression for the variables of interest at the end of the loop body. The values may contain the loop index, and the initial value ($\lambda$) as parameters. If the range expressions contain other variables, for which range expressions are known, these expressions will be substituted. If the values are not known, the range expression  becomes unknown ($\bot$).

For example, at the end of the loop on lines 3--8, of Figure~\ref{fig:CGcode} the variable $count$ will be
\begin{quote}
    count: $[\lambda:\lambda+1]$
\end{quote}
The values for arrays $column\_number$ and $value$ will be set to $\bot$, as there is insufficient information available about their right-hand-side expressions and the index expressions are not of the form $i+k$.

We will explain Phase 1 for the loops on lines 1 and 13 in Section~\ref{subsec:example}. 

\subsection{Algorithm for Phase 2}

Phase 2 aggregates the values computed by a single loop iteration across the iteration space, producing the values computed by the entire loop. This can be straightforward in the case where the value range expression of the variable contains neither the loop index nor $\lambda$,  and, in case of an array, the subscript is a simple expression. In that case, the aggregated value is the same as the one produced by a single iteration.  For array values, the loop index in the subscript expression is simply replaced by the iteration range. Consider the following example. As we will see, Phase~1  yields the expression for line 10:
\begin{quote}
    rowsize: $[i], [0:COLUMNLEN-1]$
\end{quote}
meaning a single iteration of the $i$-loop assigns a value range $[0:COLUMNLEN-1]$ to the array element $rowsize[i]$. Aggregation now expands the subscript $[i]$ to the full loop range $[0:ROWLEN-1]$. The value expression is loop invariant, hence it is the same for all array elements. The total effect of the loop is
\begin{quote}
    rowsize: $[0:ROWLEN-1], [0:COLUMNLEN-1]$
\end{quote}

For scalars, if a bound of the value range expression contains $\lambda$, the aggregated bound is computed by applying the expression (which represents the effect of one loop iteration) repeatedly $n$ times, where $n$ is the number of iterations. In the case of an expression $\lambda+k$, where $k$ is a constant, this amounts to $\Lambda+n*k$. Note that $\Lambda$  now represents the variable value at the beginning of the loop. Our current algorithm does not handle more complex bound expressions.
In the expression $[\lambda:\lambda+1]$, both bounds get processed in this way. The range expression for $count$ of the entire loop becomes $[\Lambda:\Lambda+COLUMNLEN-1]$.

The most advanced case of aggregation, needed for our example, is the recurrence relationship on line 14. Phase~1 determines:
\begin{quote}
    rowptr: $[i], rowptr[i-1] + [0:COLUMNLEN-1]$
\end{quote}

Notice that $[0:COLUMNLEN-1]$ has already been determined above to be the value range of array $rowsize$, in the subscript range $0:ROWLEN-1$. Also note that this is a non-negative number. In the case of this recurrence relationship (the array element referring back to the value written in the prior iteration) and the added term being a non-negative value, the aggregation step determines that the relationship between adjacent array elements is ``monotonically increasing", and this holds for the entire array range:
\begin{quote}
    rowptr: $[1:ROWLEN], Monotonic\_inc$
\end{quote}

There are a number of advanced cases of aggregation, which however are not needed for our initial algorithm. For example, an expression (or bound) of the form $\lambda+i$ ($i$ being the loop index, and loop bounds $0:n-1$) will yield $\Lambda+\sum_0^{n-1}i=\Lambda+n(n-1)/2$. Also, if phase~1 determines $x: [i],[i]$ (i.e., the loop body contains the array assignment $x[i]=i$), aggregation yields $x: [0:n-1], Identity$. The full aggregation algebra analyzing and expressing these cases will be presented in a forthcoming paper.

\subsection{Example}
\label{subsec:example}

The algorithm proceeds in the following sequence of phases for the code shown in Figure~\ref{fig:CGcode}:
Phase~1, followed by Phase~2 for the loops on lines 3, 1, 13. The phases yield the values:
\\[2mm]
Phase~1 (3):  $count: [\lambda:\lambda+1]; column\_number: \bot; value: \bot$\\
Phase~2 (3):  $count: [\Lambda:\Lambda+COLUMNLEN-1]; column\_number: \bot; value: \bot$
\\[2mm]
Phase~1 (1):  $count: [0:COLUMNLEN-1]; rowsize: [i], [0:COLUMNLEN-1]$\\
Phase~2 (1):  $count: [0:COLUMNLEN-1]; rowsize: [0:ROWLEN-1], [0:COLUMNLEN-1]$
\\[2mm]
Phase~1 (13):  $rowptr: [i],rowptr[i-1]+[0:COLUMNLEN-1]$\\
Phase~2 (13):  $rowptr: [1:ROWLEN],Monotonic\_inc$\\

\section{Related Work}
\label{sec:related}

The importance of such array properties as monotonicity for array subscript analysis was recognized early-on by McKinley~\cite{McKi91}.  Gutierrez et al. \cite{gutierrez2000automatic} presented run time techniques for using subscript array properties, especially monotonicity, for parallelizing sparse matrix computations. Their techniques were primarily based on pattern matching. Lin and Padua~\cite{LiPa00} took a stab at a compile-time technique to analyze the content of index arrays and automatically parallelize loops. They used a form of demand-driven, interprocedural query propagation to analyze array properties, such as injectivity, monotonicity, closed-form values, and closed-form bounds. They found several cases where they could detect closed-form values and bounds. However, they were not yet able to detect
injectivity or monotonicity, which would have been needed to parallelize the encountered loops automatically. The {\em single indexed array analysis technique}~\cite{lin2000analysis} used by Lin checks for injectivity of the index array. The method analyzes index array sections defined only in index gathering loops. We found that, in a majority of scientific applications, index arrays are filled in more general loop patterns. In~\cite{mohammadi2018extending}, Mohammadi et al.\ describe additional properties, such as Triangularity and Periodic Injectivity/Monotonicity similar to the subset property described in Section 2, and propose a framework for automatically detecting these properties, based on user-defined assertions w.r.t. the index arrays and loop dependence information. The assertions are processed by a Z3 SMT solver to test whether the constraints are satisfiable. Our work in contrast seeks to derive properties of index arrays from the program directly, using the presented algorithm.

A large number of contributions investigate array values and data dependences at runtime. Many ``inspector-executor schemes" analyze access patterns -- often via stripped-down loops that only contain the code performing the actual array accesses -- and then decide on legality or optimization of the applicable transformations. Initially, inspector-executor schemes were applied to improve communication and scheduling operations by Saltz et al.~\cite{saltz91}. More recently, they were also applied in the context of dependence analysis and automatic parallelization in Sparse Matrix Computations, such as in the work by Mohammadi et al~\cite{MYCD19} and by Venkat et al.~\cite{venkat2016automating}.  A related method is runtime data-dependence testing, which tracks references that may access the same memory location, pioneered by Rauchwerger and Padua~\cite{RaPa99}. While runtime methods can be powerful, due to the availability of complete information on access patterns at execution time, their Achilles' heel is the significant overhead of the inserted inspection and decision code. 

Work related to our analysis methods include abstract interpretation and loop aggregation methods. Abstract interpretation by Cousot and Cousot~\cite{CoCo77} symbolically executes the program and tracks values in the form of symbolic expression. A difficulty is the interpretation of loops, whose bounds are unknown in most cases. ``Widening and narrowing" has been proposed to ensure termination of the algorithms in the presence of control flow cycles. The original method is general for any control flow, but does not yield precise information about loop bounds. The Range Analysis~\cite{BlEi95} method that we use as a basis for analyzing loop bodies, uses a form of abstract interpretation. 

A technique that can produce precise information for the forms of loops that compile-time techniques usually attempt to parallelize, is the aggregation of information gathered in the loop body across the iteration space. This method was applied by Tu and Padua in array privatization~\cite{TuPa93} to analyze array sections that are defined and used. We use a similar method, extended to capture the effect of certain recurrence relationships, which allow us to gather such array properties as monotonicity. Understanding these effects is the key to extending the work by Lin and Padua, mentioned above, for successfully recognizing parallel loops.

\section{Evaluation}
\label{sec:eval}

We have applied the presented array analysis techniques to the CG code of the latest NAS Parallel Benchmarks v3.3.1 by hand. The example shown in Figure~\ref{fig:CGcode} is the key pattern in this code. The technique of Section~\ref{sec:algorithm} finds a {\em monotonically increasing} property of array $rowptr$. Based on this information, the data dependence test can determine that the iteration ranges for the loop on line 25 are non-overlapping in different iterations of the outer loop (line 18). The implementation of this data dependence test is not part of the present contribution; it is ongoing work. Key for this paper is that independence {\em can} be proven at compile time, based on  $rowptr$'s property of monotonicity. 

We briefly sketch the basic idea. We are extending the The Range Test~\cite{BlEi94b}, which symbolically computes the array ranges being accessed in the iterations of the loop being analyzed, and then tests if these ranges overlap. If they do not, the iterations are independent and thus the loop can be parallelized. For the loop on line 18, the Range Test finds that the inner loop (line 25) accesses the array range $product\_array[rowptr[i-1]:rowptr[i]]$. The key step in which the extended Range Test looks for overlap of the array ranges is a comparison of the range accessed in an arbitrary iteration $i$ and the successor iteration $i+1$. This comes down to the comparison of the two range expressions $rowptr[i-1]:rowptr[i]$ and $rowptr[i]:rowptr[i+1]$. The monotonicity property shows that $rowptr[i]<=rowptr[i+1]$ for all values of $i$ in the outer loop. Thus, the array ranges being accessed do not overlap. The reader may notice a subtlety: the first iteration is a special case, due to the if statement on lines 19--24. This case could be handled by peeling the first iteration. However, symbolic analysis can do better and prove  that there is no overlap even in the first iteration. This will be described in a forthcoming paper on the implemented Range Test extension. 

\subsection{Performance Results}
\begin{figure}[H]
\centering\includegraphics[width=4.5in]{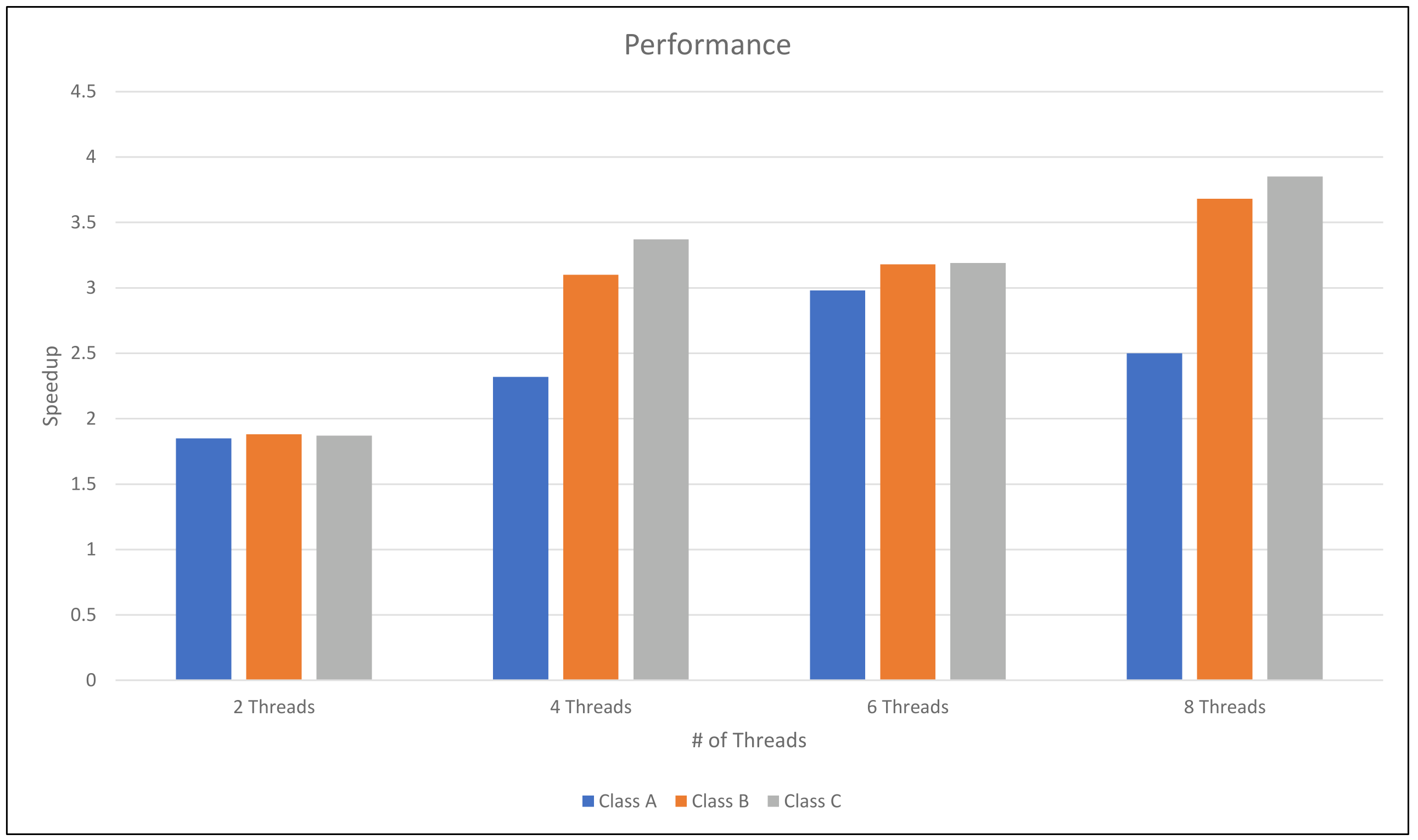}
\caption{Performance results of the CG benchmark ( NPB 3.3 ).} 
\label{fig:results_table}
\end{figure}

Figure~\ref{fig:results_table} shows the performance results of the CG benchmark, which is essentially an Unstructured Sparse Linear System Solver using the Conjugate Gradient Method \cite{bailey2011parallel}.  The benchmark includes several data sets, of increasing size, referred to as Class A, B, C etc. The speedups for the 3 Classes using 2, 4, 6 and 8 threads, have been plotted. The speedup in this case, is the performance improvement achieved after parallelizing loops with subscripted subscript patterns only, similar to the ones analyzed in Figure~\ref{fig:CGcode}, relative to the original, sequential execution.  The execution times were recorded on a machine with an Intel Kaby Lake R processor having 4 cores and 8 threads at 1.60GHz base frequency along with 8GB of 1866 MHz DDR4 memory . We used gcc v7.3.0 on Ubuntu Linux v18.04 to compile and run the application codes.

For Class A, as the number of threads increases, the performance also increases up to 6 threads. For 8 threads, the performance is only slightly better than with 4 threads. On the other hand, for Classes B and  C, the performance gain is the highest with 8 threads. We attribute this behavior to the availability of additional hardware resources for executing more than one thread per core, which benefits large data sets but not the smaller ones.  The key evaluation result, for this paper,  is that the parallel execution of loops with subscripted subscripts, which our new techniques enable, gains significant performance. Current parallelizers do not detect these loops as parallel, executing bulk of the program sequentially.


\section{Conclusions}
\label{sec:conclude}
We have presented a novel compile-time analysis method for subscripted subscripts, which can gather sufficient information to successfully parallelize an important class of programs exhibiting sparse matrix patterns. The method finds that certain arrays that are used in subscripts of other arrays are monotonic. This property provides sufficient information to data-dependence analysis, allowing it to detect that the enclosing loops are parallel. The CG code of the latest version of the NAS parallel benchmark represents this class of patterns. Applying the techniques by hand to this code, yields a parallel program that gains a speedup of 3.8 on four cores, demonstrating that the subscript patterns show up in key program sections. Current compiler technology is unable to improve this benchmark significantly.  Our techniques are the first to fully automatically parallelize certain loops containing subscripted subscripts. These patterns, in turn, are the key obstacle preventing current compiler technology from matching the hand-coded parallelism in the NAS parallel benchmark suite.

\bibliographystyle{IEEEtranS}
\bibliography{lcpc19}

\begin{thebibliography}{10}
\providecommand{\url}[1]{#1}
\csname url@samestyle\endcsname
\providecommand{\newblock}{\relax}
\providecommand{\bibinfo}[2]{#2}
\providecommand{\BIBentrySTDinterwordspacing}{\spaceskip=0pt\relax}
\providecommand{\BIBentryALTinterwordstretchfactor}{4}
\providecommand{\BIBentryALTinterwordspacing}{\spaceskip=\fontdimen2\font plus
\BIBentryALTinterwordstretchfactor\fontdimen3\font minus
  \fontdimen4\font\relax}
\providecommand{\BIBforeignlanguage}[2]{{%
\expandafter\ifx\csname l@#1\endcsname\relax
\typeout{** WARNING: IEEEtranS.bst: No hyphenation pattern has been}%
\typeout{** loaded for the language `#1'. Using the pattern for}%
\typeout{** the default language instead.}%
\else
\language=\csname l@#1\endcsname
\fi
#2}}
\providecommand{\BIBdecl}{\relax}
\BIBdecl

\bibitem{BMLA12}
\BIBentryALTinterwordspacing
H.~Bae, D.~Mustafa, J.-W. Lee, Aurangzeb, H.~Lin, C.~Dave, R.~Eigenmann, and
  S.~Midkiff, ``The cetus source-to-source compiler infrastructure: Overview
  and evaluation,'' \emph{International Journal of Parallel Programming}, pp.
  1--15, 2012, 10.1007/s10766-012-0211-z. [Online]. Available:
  \url{http://dx.doi.org/10.1007/s10766-012-0211-z}
\BIBentrySTDinterwordspacing

\bibitem{bailey2011parallel}
D.~H. Bailey, ``Nas parallel benchmarks,'' \emph{Encyclopedia of Parallel
  Computing}, pp. 1254--1259, 2011.

\bibitem{BlEi94b}
W.~Blume and R.~Eigenmann, ``{The Range Test: A Dependence Test for Symbolic,
  Non-linear Expressions},'' \emph{Proceedings of Supercomputing '94,
  Washington D.C.}, pp. 528--537, Nov. 1994.

\bibitem{BlEi95}
\BIBentryALTinterwordspacing
------, ``Symbolic range propagation,'' in \emph{the 9th International Parallel
  Processing Symposium}, 1995, pp. 357--363. [Online]. Available:
  \url{citeseer.nj.nec.com/blume95symbolic.html}
\BIBentrySTDinterwordspacing

\bibitem{CoCo77}
P.~Cousot and R.~Cousot, ``Abstract interpretation: A unified lattice model for
  static analysis of programs by construction or approximation of fixpoints,''
  in \emph{Proceedings of 4th ACM Symposium}, 1977, pp. 238--252.

\bibitem{doi:10.1137/1.9780898718881}
\BIBentryALTinterwordspacing
T.~A. Davis, \emph{Direct Methods for Sparse Linear Systems}.\hskip 1em plus
  0.5em minus 0.4em\relax Society for Industrial and Applied Mathematics, 2006.
  [Online]. Available:
  \url{https://epubs.siam.org/doi/abs/10.1137/1.9780898718881}
\BIBentrySTDinterwordspacing

\bibitem{gutierrez2000automatic}
E.~Guti{\'e}rrez, R.~Asenjo, O.~Plata, and E.~L. Zapata, ``Automatic
  parallelization of irregular applications,'' \emph{Parallel Computing},
  vol.~26, no. 13-14, pp. 1709--1738, 2000.

\bibitem{IntelICC}
Intel. (2011) Automatic parallelization with intel compilers. Intel.
  \url{https://software.intel.com/en-us/articles/automatic-parallelization-with-intel-compilers}
  , visited 10-21-2019.

\bibitem{lin2000analysis}
Y.~Lin and D.~Padua, ``Analysis of irregular single-indexed array accesses and
  its applications in compiler optimizations,'' in \emph{International
  Conference on Compiler Construction}.\hskip 1em plus 0.5em minus 0.4em\relax
  Springer, 2000, pp. 202--218.

\bibitem{LiPa00}
\BIBentryALTinterwordspacing
------, ``Compiler analysis of irregular memory accesses,'' in
  \emph{Proceedings of the ACM SIGPLAN 2000 Conference on Programming Language
  Design and Implementation}, ser. PLDI '00.\hskip 1em plus 0.5em minus
  0.4em\relax New York, NY, USA: ACM, 2000, pp. 157--168. [Online]. Available:
  \url{http://doi.acm.org/10.1145/349299.349322}
\BIBentrySTDinterwordspacing

\bibitem{McKi91}
K.~S. McKinley, ``{Dependence Analysis of Arrays Subscripted by Index
  Arrays},'' Rice University, Tech. Rep., June 1991, tR91-162.

\bibitem{mohammadi2018extending}
M.~S. Mohammadi, K.~Cheshmi, M.~M. Dehnavi, A.~Venkat, T.~Yuki, and M.~M.
  Strout, ``Extending index-array properties for data dependence analysis,''
  2018.

\bibitem{MYCD19}
\BIBentryALTinterwordspacing
M.~S. Mohammadi, T.~Yuki, K.~Cheshmi, E.~C. Davis, M.~Hall, M.~M. Dehnavi,
  P.~Nandy, C.~Olschanowsky, A.~Venkat, and M.~M. Strout, ``Sparse computation
  data dependence simplification for efficient compiler-generated inspectors,''
  in \emph{Proceedings of the 40th ACM SIGPLAN Conference on Programming
  Language Design and Implementation}, ser. PLDI 2019.\hskip 1em plus 0.5em
  minus 0.4em\relax New York, NY, USA: ACM, 2019, pp. 594--609. [Online].
  Available: \url{http://doi.acm.org/10.1145/3314221.3314646}
\BIBentrySTDinterwordspacing

\bibitem{PGI2018}
\BIBentryALTinterwordspacing
PGI. (2018) Pgi compiler user's guide. Nvidia. [Online]. Available:
  \url{https://www.pgroup.com/resources/docs/18.4/openpower/pgi-user-guide/index.htm}
\BIBentrySTDinterwordspacing

\bibitem{quinlan2011rose}
D.~Quinlan and C.~Liao, ``The rose source-to-source compiler infrastructure,''
  in \emph{Cetus users and compiler infrastructure workshop, in conjunction
  with PACT}, vol. 2011.\hskip 1em plus 0.5em minus 0.4em\relax Citeseer, 2011,
  p.~1.

\bibitem{RaPa99}
\BIBentryALTinterwordspacing
L.~Rauchwerger and D.~A. Padua, ``The {LRPD} test: Speculative run-time
  parallelization of loops with privatization and reduction parallelization,''
  \emph{IEEE Transactions on Parallel and Distributed Systems}, vol.~10, no.~2,
  pp. 160--??, 1999. [Online]. Available:
  \url{citeseer.nj.nec.com/rauchwerger95lrpd.html}
\BIBentrySTDinterwordspacing

\bibitem{saltz91}
J.~Saltz, R.~Mirchandaney, and K.~Crowley, ``Run time parallelization and
  scheduling of loops.'' \emph{IEEE Transactions on Computers}, vol.~40, no.~5,
  pp. 603--612, May 1991.

\bibitem{TuPa93}
P.~Tu and D.~Padua, ``{Automatic Array Privatization},'' in \emph{Proc. Sixth
  Workshop on Languages and Compilers for Parallel Computing, Portland, OR.
  Lecture Notes in Computer Science.}, U.~Banerjee, D.~Gelernter, A.~Nicolau,
  and D.~Padua, Eds., vol. 768, August 12-14, 1993, pp. 500--521.

\bibitem{venkat2016automating}
A.~Venkat, M.~S. Mohammadi, J.~Park, H.~Rong, R.~Barik, M.~M. Strout, and
  M.~Hall, ``Automating wavefront parallelization for sparse matrix
  computations,'' in \emph{Proceedings of the International Conference for High
  Performance Computing, Networking, Storage and Analysis}.\hskip 1em plus
  0.5em minus 0.4em\relax IEEE Press, 2016, p.~41.

\end{thebibliography}

\end{document}